\documentclass{IOS-Book-Article}
\usepackage[utf8x]{inputenc}
\usepackage{comment}
\usepackage{mathptmx}
\usepackage{graphicx}
\usepackage{subcaption}
\usepackage{paralist}
\usepackage{amsmath}                    % For equation environment
\usepackage{amssymb}
\usepackage{hyperref}                 % For inserting PDF options (author etc.)
\usepackage[usenames,dvipsnames]{xcolor} % Colored text
\usepackage{eurosym}
\usepackage{multirow}
\definecolor{darkred}{rgb}{0.5,0,0}
\definecolor{darkgreen}{rgb}{0,0.5,0}
\definecolor{darkblue}{rgb}{0,0,0.5}
\definecolor{gray}{rgb}{0.35,0.35,0.35}
\definecolor{lgray}{rgb}{0.9,0.9,0.9}

\usepackage{listings}                   % For code formatting
\newcommand{\mset}[1]{\left\{#1\right\}}
\newcommand{\mpar}[1]{\left(#1\right)}

\newcommand{\oper}[5]{\genfrac{}{}{}{0}{#1 \xrightarrow{#2} #3}{#4 \to #5}}

\newcommand{\operdsplit}[5]{\genfrac{}{}{}{0}{#1 \xrightarrow{#2} #3}{
\begin{array}{c}
    #4 \to\\
    \hspace{2pt}#5
\end{array}}}

\newcommand{\operndsplit}[5]{\genfrac{}{}{}{0}{
\begin{array}{c}
    #1 \xrightarrow{#2}\\
    \hspace{2pt}#3
\end{array}
}{
\begin{array}{c}
    #4 \to\\
    \hspace{2pt}#5
\end{array}}}

\def\hb{\hbox to 10.7 cm{}}

\usepackage[textsize=tiny,textwidth=3cm]{todonotes}
\makeatletter
\renewcommand{\@todonotes@drawLineToRightMargin}{%
\if@todonotes@line%
\if@todonotes@fancyline%
\tikz[remember picture,overlay]{%
\tikzstyle{both}=[line width=1pt, color=\@todonotes@currentbackgroundcolor, draw, opacity=0.35]%
\tikzstyle{line}=[shorten >=0pt, line cap=round]%
\tikzstyle{head}=[shorten >=-1pt, dash pattern=on 0pt off 1pt, ->]%
\foreach \s in {line,head}{%
\draw[both,\s]%
(inNote.180).. controls +(180:2) and +(30:1.5)..([xshift=2pt]inText);%
(inNote.180).. controls +(180:2) and +(30:1.5)..([xshift=2pt]inText);%
};%
}%
\else%
\begin{tikzpicture}[remember picture, overlay]%
\draw[connectstyle]%
([yshift=-0.2cm] inText)%
-| ([xshift=-0.2cm] inNote.west)%
-| (inNote.west);%
\end{tikzpicture}%
\fi%
\fi}%
\makeatother

\begin{document}

\pagestyle{plain} % JAP 2019-05-28 so we can see how far we have to go!

\begin{frontmatter}              % The preamble begins here.

%\pretitle{Pretitle}
\title{Towards a Goal-oriented Agent-based Simulation framework for High-Performance Computing}

\markboth{}{May 2019\hb}
%\subtitle{Subtitle}

\author[A]{\fnms{Dmitry} \snm{Gnatyshak}},
\author[A]{\fnms{Luis} \snm{Oliva-Felipe}},
\author[B]{\fnms{Sergio} \snm{\'Alvarez-Napagao}},
\author[C]{\fnms{Julian} \snm{Padget}},
\author[A]{\fnms{Javier} \snm{V\'azquez-Salceda}},
\author[B]{\fnms{Dario} \snm{Garcia-Gasulla}}
and
\author[{A,B}]{\fnms{Ulises} \snm{Cort\'es}}
\runningauthor{Dmitry Gnatyshak}
\address[A]{Universitat Polit\`ecnica de Catalunya - BarcelonaTECH (UPC)\\
				              C/Jordi Girona 1-3, E-08034 , Barcelona, Spain}
\address[B]{Barcelona Supercomputer Centre (BSC)\\
				              C/Jordi Girona 1-3, E-08034 , Barcelona, Spain}
\address[C]{Department of Computer Science\\
				              University of Bath, BATH BA2 7AY , United Kingdom}
\begin{abstract}
Currently, agent-based simulation frameworks force the user to choose between simulations involving a large number of agents (at the expense of limited agent reasoning capability) or simulations including agents with increased reasoning capabilities (at the expense of a limited number of agents per simulation). This paper describes a first attempt at putting goal-oriented agents into large agent-based (micro-)simulations. We discuss a model for goal-oriented agents in High-Performance Computing (HPC) and then briefly discuss its implementation in PyCOMPSs (a library that eases the parallelisation of tasks) to build such a platform that benefits from a large number of agents with the capacity to execute complex cognitive agents.
\end{abstract}

\end{frontmatter}
\markboth{May 2019\hb}{May 2019\hb}

\section{Introduction}
Agent-based simulation (ABS) faces the conflicting demands of providing larger simulations (in terms of the number of agents) and providing agents with better cognitive and decision-making capabilities (which tend not to scale well in large simulations).
%in both memory demands and execution time). 
This paper is a first attempt at addressing this friction, through the use of High-Performance Computing (HPC), with the aim of % bringing goal-oriented agents into agent-based simulations with large numbers of individuals. This allows for the 
enabling scenarios in which large populations of individuals (e.g. traffic simulation, industrial and urban areas along geographical areas) have the ability to perform normative reasoning, planning or even BDI-like behaviour, in order to explore the analysis of phenomena resulting from more detailed behavioural modelling. On the one hand, current HPC-based approaches\cite{Rousset2016} seem to focus on large simulations with limited interaction or perception or with reactive-like agents. In some cases, planning is even hard-coded. On the other hand, \cite{Abar2017} lists a wide range of ABS which seem to hardly offer scalability and agents with high cognitive capabilities. This is an indication of an existing trade-off between scalability and agents with complex deliberation processes.
In Section 2, we discuss a model for goal-oriented agents in HPC and briefly outline its implementation in PyCOMPSs (a library that eases task parallelisation through annotation) to build such a platform. %that benefits from a large number of agents with the capacity to execute complex cognitive agents.
In Section 3, we present the test scenarios used to evaluate the performance of the platform and demonstrate its capabilities and potential. %This scenario is based on a river basin in which industries and wastewater treatment plants interact with the overall goal of maintaining a sustainable (law-compliant) river water quality level with respect to pollution.
In Section 4, we summarise the main results, the contributions and we identify the next steps.

\section{Proposed Model}
This section provides background about the COMPSs HPC framework and the corresponding Python package, and then presents the proposed agent-based simulation model.

\subsection{COMPSs and PyCOMPSs}

COMPSs \cite{Tejedor2008} is a framework based on the Grid Component Model (GCM)\cite{gcm2007} 
and ServiceSs model\cite{Tejedor2011} developed by the Barcelona Supercomputing Center.
Its main purpose is to allow the development of distributed cloud- or grid-based applications without the need to deal with the specifics of underlying execution systems.
Thus, it provides an abstraction layer, allowing the development of hardware-configuration-agnostic applications. These applications can be distributed automatically, saving effort in accounting for low-level aspects of the target hardware.
COMPSs analyses the data dependencies among the user-specified functions (called \textit{tasks} as in the GCM model) of a sequential program and runs as many of those tasks in parallel as it is safe to do.
A set of additional commands and parameters can be used to fine-tune the runtime execution and parallelisation.
The core of this runtime is written in Java, but C/C++ and Python interfaces are also available.
As all of these commands are language-specific, for the purpose of this work we choose to focus on the Python
version: PyCOMPSs\cite{pycompss2017}.

%From the runtime point of view, PyCOMPSs adds a few layers to the basic COMPSs Java runtime, as shown in Fig.\ref{fig:pycompss1}.
%Essentially, what these additional layers do is transform the Python data types into Java ones (or into files for user classes), for proper data dependency analysis by the main runtime.
%Python bindings gather these types of information by analyzing the code and using the auxiliary C++ library for proper binding and transformation.
%Note, however, that this approach adds an additional requirement: the custom class objects in the Python code must be serializable.

\begin{figure}[!t]
    \begin{subfigure}[b]{0.455\textwidth}
    \includegraphics[width=0.95\textwidth]{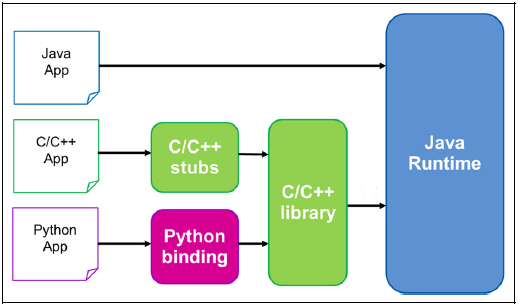}
    \caption{PyCOMPSs runtime structure}
    \label{fig:pycompss1}
    \end{subfigure}\hfill%
    \begin{subfigure}[b]{0.545\textwidth}
    \includegraphics[width=0.95\textwidth]{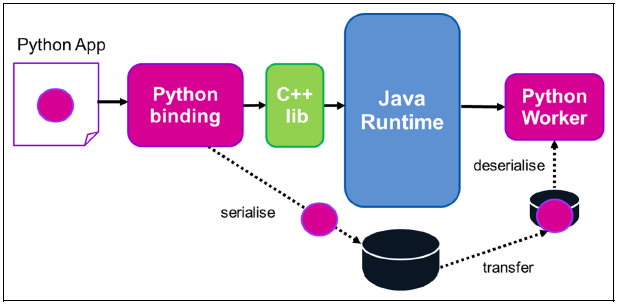}
    \caption{PyCOMPSs object handling scheme}
    \label{fig:pycompss2}
    \end{subfigure}
    \caption{Aspects of PYCOMPSs}
\end{figure}

From a coding perspective, the %changes to the original are minimal. The
COMPSs interface takes the form of annotations (decorators) in the Python code and several API functions and procedures.
The main annotation is \texttt{@Task(\dots)}, where the user specifies the return type (a type or class name) and the data direction of inputs (in, out) for mutable types and classes.%(all immutable types are naturally read-only, the direction is IN, and by default all the inputs are also IN, so it does not need to be specified).
The following code shows an example of a PyCOMPSs task annotation:
\begin{lstlisting}
@Task(par1=INOUT, par3=OUT, returns=list)
def my_func(par1, par2, par3):
\end{lstlisting}
% \begin{flushleft}
% \colorbox{lgray}{\texttt{@Task(\pycom{par1}=INOUT, \pycom{par3}=OUT, \pycom{returns}=\pytype{list})}}\\
% \colorbox{lgray}{\texttt{\pycom{def} my\_func(par1, par2, par3):}}
% \end{flushleft}
By means of such annotations, we may mark global functions, class methods, and instance methods as tasks to be automatically parallelised.

\subsection{Conceptual Model: Multi-agent system  elements}\label{section:model-mas}

%In this section we introduce the formal conceptual model of our agent-based HPC simulation framework. The main purpose is to present the main elements of the framework, and then to formally describe their operational semantics in sections 2.3 and 2.4. 

We define a {\bf Multi-agent System} $\mathcal{M}$ as the tuple %:
%\begin{align*}
   $\mathcal{M}=\mset{E, \mathcal{A}^+, \mathcal{C}} $
%\end{align*}
where:
\begin{compactenum}
    \item $E$ is an {\bf environment}, in which the agents are situated, that they can perceive, and on which they can act
    \item $\mathcal{A}^+$ is a non-empty set of agents
    \item $\mathcal{C}$ is a controller that maintains the environment and handles communications between agents
\end{compactenum}
An {\bf Agent} $\mathcal{A}_i$ is defined as %:
%\begin{align*}
    $\mathcal{A}_i=\mset{ID, msgQs, Bh, \mathbb{B}, \mathbb{G}, \mathcal{P}, outAcs}$
%\end{align*}
where:
\begin{compactenum}
    \item $ID=\mset{AgID, AgDesc}$ is $\mathcal{A}_i$'s identity data
    \begin{compactenum}
        \item $AgID$ is the unique identifier for $\mathcal{A}_i$
        \item $AgDesc$ is an arbitrary description of $\mathcal{A}_i$
    \end{compactenum}
    \item $msgQs=\mset{\mathcal{I}, \mathcal{O}}$ is a set of the $\mathcal{A}_i$'s message queues
    \begin{compactenum}
        \item $\mathcal{I}=\mset{\dots,msg_i,\dots}$ is the inbox, a set of messages sent to $\mathcal{A}_i$
        \item $\mathcal{O}=\mset{\dots,msg_i,\dots}$ is the outbox, a set of messages sent by $\mathcal{A}_i$
        \item $msg_i=\mset{AgID_s, AgID_r, performative, content, priority}$ is a {\bf message} from agent $Ag_s$ to agent $Ag_r$ with the given performative, content, and priority. $performatives$ are FIPA-compliant.
    \end{compactenum}
    \item $Bh=\mset{MendR, \mathbb{RG}}$ is $\mathcal{A}_i$'s {\bf role behaviour}
    \begin{compactenum}
        \item $MendR$ is a {\bf means-ends reasoner} used to generate plans (see Section 2.5)
        \item $\mathbb{RG}$ is a set of goals associated with this role behaviour
    \end{compactenum}
    \item $\mathbb{B}$ is the set of $\mathcal{A}_i$'s {\bf beliefs}
    \item $\mathbb{G}$ is the set of $\mathcal{A}_i$'s {\bf goals}
    \item $\mathcal{P}=\mset{\dots,ab_i,\dots}$ is the {\bf current plan} of $\mathcal{A}_i$
    \item $ab_i=\mset{\dots,a_{ij},\dots}$ is an {\bf action block}, an ordered set of agent actions, for which we distinguish three types:
    %\begin{compactdesc}
        %\item[Internal actions:] 
        {\bf internal actions} executed by the agents and intended to change their beliefs
        %\item[External actions:] 
        {\bf external actions} sent to the controller to be executed by it on the environment, and
        %\item[Message actions:] 
        {\bf message actions} to generate messages to other agents
    %\end{compactdesc}
    \item $outAcs$ is a set of external actions to be executed on the environment
    \begin{compactenum}
        \item $outAcs_i=\mset{senderID, a^e}$ is a tuple of sender ID and an external action
    \end{compactenum}
\end{compactenum}

The {\bf Controller} $\mathcal{C}$ is defined by the tuple %:
%\begin{align*}
$\mathcal{C}=\mset{\mathcal{I},inAcs}$
%\end{align*}
where $\mathcal{I}$ is an inbox for the reception of messages from all the agents and $inAcs$ is the set of all the actions to be applied on the environment.

\subsection{Operational semantics of the Agents' deliberation cycle: agent transition rules}\label{subsec:model-agents}
In the previous section we have defined the different elements that compose our framework. Using these definitions, we can introduce the operational semantics of our model by means of a set of transition rules.
The model assumes that $\mathcal{M}$ evolves as a set of simulation steps.
These transition rules describe how the agents' internal states are transformed over each single step of the simulation.
They show the part of the state that is changed, the function that changes it, and a summary of the transformation.

Agents in our model are goal-oriented deliberative agents inspired by the Beliefs-Desires-Intentions (BDI) Model\cite{rao91a}. Therefore, the agents' reasoning cycle is split into a \emph{perception} phase (where beliefs are updated), a \emph{deliberation} phase (where goals are updated and prioritised), a \emph{means-ends reasoning} phase (where plans are selected/built) and an \emph{actuation} phase. As in many modern BDI-inspired agent languages (e.g., JADEX, 2APL, GOAL), in our model plans are selected directly from beliefs and goals without the need for intermediate intention selection (which forces, in the theoretical BDI model, to only pursue one goal at a time).

The first sub-step in the agent deliberation cycle happens when the agent perceives the current state of the environment.
This perception only modifies the agent's beliefs which are then used in the following sub-steps.
Formally, this sub-step is formalised in Rule~\ref{eq:perceive} (formulae are provided in Figure~\ref{fig:agentrules}):

\begin{figure}[!t]
  \begin{centering}
    \small
    \begin{tabular}{|p{5.6cm}|p{5.6cm}|}
      \hline 
      \begin{equation}
        \label{eq:perceive}
        \operdsplit{\mathbb{B}}{perceive\mpar{E}}{\mathbb{B}^\prime}
        {\mathcal{A}_i\mset{\cdot,\cdot,\cdot,\mathbb{B},\cdot,\cdot,\cdot},E}
        {\mathcal{A}_i\mset{\cdot,\cdot,\cdot,\mathbb{B}^\prime,\cdot,\cdot,\cdot},E}
      \end{equation}
      &
        \begin{equation}
          \label{eq:message}
          \operdsplit{\mathbb{B},\mathbb{G},\mathcal{O}}{process\mpar{h}}{\mathbb{B}^\prime,\mathbb{G}^\prime,\mathcal{O}^\prime}
          {\mathcal{A}_i\mset{\cdot,\mset{\mset{h,t},\mathcal{O}},\cdot,\mathbb{B},\mathbb{G},\cdot,\cdot},\cdot}
          {\mathcal{A}_i\mset{\cdot,\mset{\mset{t},\mathcal{O}^\prime},\cdot,\mathbb{B}^\prime,\mathbb{G}^\prime,\cdot,\cdot},\cdot}
        \end{equation}
      \\ \hline
      \begin{equation}
        \label{eq:goal}
        \operdsplit{\mathbb{G}}{goal\_check\mpar{\mathbb{B},\mathbb{G}}}{\mathbb{G}^\prime}{\mathcal{A}_i\mset{\cdot,\cdot,\cdot,\cdot,\mathbb{G},\cdot,\cdot},\cdot}{\mathcal{A}_i\mset{\cdot,\cdot,\cdot,\cdot,\mathbb{G}^\prime,\cdot,\cdot},\cdot}
      \end{equation}
      &
        \begin{equation}
          \label{eq:meansends}
          \operdsplit{\mathcal{P}=\varnothing\vee \mathcal{P}=Fail}{MendR\mpar{\mathbb{B},\mathbb{G},\mathbb{RG}}}{\mathcal{P}^\prime}{\mathcal{A}_i\mset{\cdot,\cdot,\cdot,\cdot,\cdot,\mathcal{P},\cdot},\cdot}{\mathcal{A}_i\mset{\cdot,\cdot,\cdot,\cdot,\cdot,\mathcal{P}^\prime,\cdot},\cdot}
        \end{equation}
      \\ \hline
      \begin{equation}
        \label{eq:action}
        \operdsplit{\mathcal{P}\mpar{h,t}}{execute\mpar{h}}{\mathcal{P}^\prime\mpar{t}}{\mathcal{A}_i\mset{\cdot,\mset{\mathcal{I},\mathcal{O}},\cdot,\mathbb{B},\cdot,\mathcal{P},outAcs},\cdot}{\mathcal{A}_i\mset{\cdot,\mpar{\mathcal{I},\mathcal{O}^\prime},\cdot,\mathbb{B}^\prime,\cdot,\mathcal{P}^\prime,outAcs^\prime},\cdot}
      \end{equation}
      &
        \begin{equation}
          \label{eq:role}
          \operdsplit{Bh}{role\_check\mpar{\mathbb{B},\mathbb{G},\mathbb{RG}}}{Bh^\prime}{\mathcal{A}_i\mset{\cdot,\cdot,Bh,\cdot,\cdot,\cdot,\cdot},\cdot}{\mathcal{A}_i\mset{\cdot,\cdot,Bh^\prime,\cdot,\cdot,\cdot,\cdot},\cdot}
        \end{equation}
      \\ \hline
    \end{tabular}
  \end{centering}
  \caption{Definitions for the agent transition (rules 1 to 6). We use $\cdot$ to denote elements in the agent tuple that are not modified in the transition.}
  \label{fig:agentrules}
\end{figure}

\paragraph{Rule~\ref{eq:perceive}:} Perceiving the environment, where $perceive\mpar{\dots}$ is a user-defined function that transforms the agent $\mathcal{A}$'s beliefs based on the environment state $E$.
%We use $\cdot$ to denote elements in the agent tuple that are not modified.

During the second sub-step, the agent processes the incoming messages.
This is done sequentially, in an arbitrary order.
Message processing may modify the agent's beliefs, as well as its goals.
This is formalised in Rule~\ref{eq:message}:

\paragraph{Rule~\ref{eq:message}} Message processing, where $process\mpar{\dots}$ is a user-defined message processing function that can modify the agent $\mathcal{A}$'s beliefs, goals, and outbox.

In the third sub-step of the deliberation cycle, the agent gets an opportunity to reevaluate and change its own goals.
The decision is based on its beliefs and current goals, and only the latter is changed during the process. This transition is defined in Rule~\ref{eq:goal}:

\paragraph{Rule~\ref{eq:goal}} Goal check, where $goal\_check\mpar{\dots}$ is a user-defined function that may change the agent $\mathcal{A}$'s goals based on its beliefs.

One of the most important sub-steps is the actual reasoning.
In this step, the agent relies on a means-ends reasoner to generate a plan that will be followed by itself afterwards.
This step is only performed if the current plan has failed or finished.\footnote{See \S\,\ref{section:model-reasoning} for more details.}
Formally, this step is defined in Rule~\ref{eq:meansends}:

\paragraph{Rule~\ref{eq:meansends}} Means-ends reasoner, where $MendR\mpar{\dots}$ is a means-ends reasoner that generates plans based on the agent $\mathcal{A}$'s beliefs and goals.

After the plan has been updated, the agent proceeds to execute it.
The next action block is extracted from the plan and the set of its actions are run one after another.

\paragraph{Rule~\ref{eq:action}} Action execution, where $execute\mpar{\dots}$ is the action execution function for action $h$. If $h$ is an 
%\begin{compactdesc}
    %\item[Internal action:]
    internal action, $execute$ runs $h$ to modify the set of agent $\mathcal{A}$'s beliefs; 
    if $h$ is an external action, 
    %\item[External action:] 
    $execute$ appends $h$ to the set of outgoing external actions;
    if $h$ is a message action,
    %\item[Message action:] 
    $execute$ runs $h$ to generate messages to send.
%\end{compactdesc}

Finally, the agent has an optional choice to reevaluate its role behaviour (i.e. $Bh$).
This sub-step is similar to the goal check and its result can be either a new role, or no changes.

\paragraph{Rule~\ref{eq:role}} Role check (optional), where $role\_check\mpar{\dots}$ is a user-defined function that may modify the agent $\mathcal{A}$'s role behaviour and role goals, based on its beliefs and goals.

\subsection{Operational semantics of Controller and Environment: MAS transition rules}\label{subsec:model-mas}
After formally defining the internal deliberation process of agents in our framework, we now describe how the Controller changes the environment on each simulation step.

First of all, there is an optional sub-step during which the controller can modify the environment before agents' actions are applied to it. These actions are considered to be commutative. Formally, we can express it in the following way:

\begin{figure}[!t]
  \begin{centering}
    \begin{tabular}{|p{5.6cm}|p{5.6cm}|}
      \hline
      \begin{equation}
        \label{eq:preaction}
        \oper{E}{pre\_step\mpar{}}{E^\prime}{\mathcal{M}\mset{E,\cdot,\cdot}}{\mathcal{M}\mset{E^\prime,\cdot,\cdot}}
      \end{equation} &
        \begin{equation}
          \label{eq:forwarding}
          \operndsplit{\mathcal{A}_r\mset{\cdot,\mset{\mset{x},\mathcal{O}},\cdot,\cdot,\cdot,\cdot,\cdot}}{fwd\_msg\mpar{h,\mathcal{A}_r}}{\mathcal{A}_r^\prime\mset{\cdot,\mset{\mset{x,h},\mathcal{O}},\cdot,\cdot,\cdot,\cdot,\cdot}}{\mathcal{M}\mset{\cdot,\mset{\dots,\mathcal{A}_r,\dots},\mset{\mset{h,t},\cdot}}}{\mathcal{M}\mset{\cdot,\mset{\dots,\mathcal{A}_r^\prime,\dots},\mset{\mset{t},\cdot}}}
        \end{equation} \\ \hline
      \begin{equation}
        \label{eq:execution}
        \operdsplit{E}{execute\_ac\mpar{a}}{E^\prime}{\mathcal{M}\mset{E,\cdot,\mset{\cdot,\mset{a,t}}}} {\mathcal{M}\mset{E^\prime,\cdot,\mset{\cdot,\mset{t}}}}
      \end{equation} &
      \begin{equation}
        \label{eq:postaction}
        \oper{E}{post\_step\mpar{}}{E^\prime}{\mathcal{M}\mset{E,\cdot,\cdot}} {\mathcal{M}\mset{E^\prime,\cdot,\cdot}}
      \end{equation} \\ \hline
    \end{tabular}
  \end{centering}
  \caption{Definitions for the MAS transition rules (rules 7 to 10)}
  \label{fig:masrules}
\end{figure}

\paragraph{Rule~\ref{eq:preaction}} Pre-action execution step (optional), where $pre\_step\mpar{}$ is a user-defined function that modifies the environment before the agents' actions are collectively applied to it.

After that, the controller acts as a message dispatcher, processing all the outgoing messages and passing them to the corresponding recipients in a sequential manner.

\paragraph{Rule~\ref{eq:forwarding}} Message forwarding, where $fwd_msh\mpar{\dots}$ is a function that moves the specified message from the sender's inbox, to the recipient's outbox.

The next sub-step executes, sequentially, all the actions sent by agents.
%This process is also done sequentially.

\paragraph{Rule~\ref{eq:execution}} Action execution, where $execute\_ac\mpar{\dots}$ is a function that applies the action $h$ to the environment.

Finally, there is another optional sub-step to modify the environment.
It is similar to the first sub-step, differing only in the function used.

\paragraph{Rule~\ref{eq:postaction}} Post-action execution step (optional), where $post\_step\mpar{}$ is a user-defined function that modifies the environment after the agents' actions are %collectively 
applied to it.

\subsection{Means-ends reasoning: Hierarchical Task Networks planner}\label{section:model-reasoning}

In our base model, we choose not to specify any fixed means-to-ends reasoning model ($MendR\mpar{\mathbb{B}, \mathbb{G}, \mathbb{RG}}$ in agent \textbf{Rule~\ref{eq:meansends}}), leaving it to the user to select one (and to provide the required specification).
However, for our implementation of the base model, we choose a
planner based on Hierarchical Task Networks (HTN)\cite{Erol1994} %\cite{Erol2003}%
to instantiate the $MendR$ function, on the grounds that HTN planners are a good fit to the means-end stage of the BDI reasoning process\cite{Sardina2011}.%\jnote{pls review this and consider where we outline BDI: see Javi's comment in slack}
There are several HTN-based models, with different representation and planning procedures.
As most of these models deal only with high-level planning, at a distance from the practical planning and following task execution, we considered HTN planner models used for actual computer games.
These models have been adapted not only for effective abstract-to-concrete facts mapping, but also for re-planning during execution.

%As a result, 
We slightly modified \cite{HTNGameAI} to suit both our needs and the COMPSs requirements.
These modifications cover the difference in the perception models (the original algorithm used active sensors), and allow the partial sets of method sub-tasks to be accepted as parts of the plan to gain flexibility. %in the model.
In this model, abstract tasks are divided into compound tasks and methods.
A compound task consists only of an ordered set of methods while each method consists of an ordered set of compound or primitive tasks, conditions and some additional information.
Primitive tasks contain action blocks to be executed, consisting of action of various types, and also some preconditions.

To connect the output of the deliberation sub-step with the HTN planner, we consider the goals provided by the $goal\_check$ function (\textbf{rule 3}) as tasks to be searched in the HTN.
By following the model planning rules associated with the (abstract) task \cite{HTNGameAI}, we can easily obtain a sequence of actions to be executed based on the current beliefs and goals.
%
%In the current version of our HTN model, 
After the current plan has finished, we generate a new one, as per \textbf{rule~\ref{eq:meansends}}.
Thus, the only explicit goal type supported currently is the maintain goal\cite{Dastani2006}, but other types of goals (for example, achieve goals or perform goals) can be simulated by the appropriate use of the $goal\_check$ function. In the case of HTN as a means-ends reasoner, $goal\_check\mpar{}$ can easily switch between different compound tasks in HTN (or select a new HTN) as a method of goal revision.
This can be done either by traversing the graph of the HTN, or by generating a new
HTN and replacing the old one, although one should always be careful about
the belief sets and ontologies used by each planner.
%%% Local Variables:
%%% mode: latex
%%% TeX-master: "../CCIA"
%%% End:

\section{Implementation and Results}
In the previous section, we introduced the agent-based simulation model and described its design features.
This section describes an instantiation of the framework in a concrete implementation and provides some preliminary test results.
%First, we briefly describe the main points of the model implementation in section~\ref{section:impl-mas}, and then show a selection on some of the performance tests we have made to it in section~\ref{section:river-sim}.

\subsection{Implementation}\label{section:impl-mas}
The whole system is implemented in Python. It is organised as a package with several modules and a list of main classes that are easily accessible by the user:
\begin{compactdesc}
    \item[Controller:] defines the \texttt{Controller} class, the main utility entity in the system, responsible for containing the MAS, running PyCOMPSs tasks, forwarding messages, and executing all the utility methods.
    \item[Agent:] defines the \texttt{Agent} class that is a container for the agent description.
    \item[Behaviour:] defines the \texttt{Behaviour} class, that implicitly provides the agent's workflow; the user may subclass this for complex behaviours.
    \item[Directory:] defines the \texttt{Directory} and \texttt{DirectoryEntry} classes that are used by the built-in directory facilitator.
    \item[HTN:] defines all the HTN-related classes, as well as some related functionality; the full list is: \texttt{CompoundTask}, \texttt{Method}, \texttt{PrimitiveTask}, \texttt{Effect}, \texttt{Action}, \texttt{ActionBlock}, \texttt{HTNPlanner}, \texttt{BeliefSet} and \texttt{Conditions}.
    \item[Messaging:] defines the \texttt{Message} and \texttt{Messagebox} classes used for agent messaging.
    \item[State:] defines the \texttt{State} structure, used to transfer and work with persistent states.
\end{compactdesc}

The \texttt{Behaviour} class follows the model presented in Section~\ref{subsec:model-agents} with its instance functions corresponding to the functions in the rules.
In complex scenarios, it is expected that the user redefines the following functions:
\begin{lstlisting}
state.beliefs = self.perceive(environment, state.beliefs)
for message in inbox:
    state.beliefs, state.planner, reply = self.process(message, state.beliefs, state.planner)
state.planner = self.goal_check(state.beliefs, state.planner)
role = self.role_check(state.beliefs)
\end{lstlisting}

A simulation is run by initialising a Controller object, registering agents in it, and calling the \texttt{run} method.
An example of the latter two:
\begin{lstlisting}
controller.generate_agent("Test agent", behavior=MyBehavior(), beliefs=myBeliefs, init_block=None, planner=myPlanner, default_block=None, services=["testing"], register=True)
controller.run(num_iter=10, performance=False)
\end{lstlisting}

An HTN planner can be generated following the structure defined in Section~\ref{section:model-reasoning}.
The user needs to create a structure of \texttt{PrimitiveTask}s, \texttt{Method}s, and \texttt{CompoundTask}s and use the root \texttt{CompoundTask} as a parameter for the planner's initialisation.

\subsection{Performance experiments}\label{section:river-sim}

In order to test the system's performance, we have designed a number of experiment sets to test different aspects of the system. The overall time performance of the system was evaluated in accordance to 4 parameters: number of agents (100), number of requested processing units (246), number of messages sent (10), and size of messages sent (0Kb).
For each series of tests one of the parameters was modified while the others were fixed at default values (the ones inside the brackets above).
Due to space limitations we will only focus on the first two. 

The Agents' behaviours are guided by a trivial HTN with a single compound task, single method, and single primitive task with 2 actions: send messages and increment step counter belief. On each turn, each agent sent a specified number of messages containing lists of zeroes of specified length to random agents and incremented its ``step'' counter belief. Upon the reception of a message, each agent incremented its ``counter'' belief.

All of these experiments, with two exceptions, were performed on the BSC NordIII cluster using the default setting of 256 processors cores per test.
%Each core is Intel SandyBridge-EP E5–2670 at 2.6 GHz.
% \begin{figure}[h]
%     \centering
%     \includegraphics[width=0.75\textwidth]{images/res-acount.png}
%     \caption{Results for agent count test. All the results are averages of 5 replications.}
%     \label{fig:res-acount}
% \end{figure}

\begin{table}[]
\centering
\begin{tabular}{|r|r|r|}
\hline
\multicolumn{1}{|c|}{\multirow{2}{*}{\begin{tabular}[c]{@{}c@{}}Number\\ of agents\end{tabular}}} & \multicolumn{2}{c|}{Time, s}                            \\ \cline{2-3} 
\multicolumn{1}{|c|}{}                                                                            & \multicolumn{1}{c|}{Total} & \multicolumn{1}{c|}{Tasks} \\ \hline
10                                                                                                & 29.5460                    & 28.5788                    \\
50                                                                                                & 75.2455                    & 72.3717                    \\
100                                                                                               & 132.4631                   & 128.8426                   \\
200                                                                                               & 308.3213                   & 302.9179                   \\
300                                                                                               & 539.4989                   & 530.3990                   \\
500                                                                                               & 972.8603                   & 959.3281                   \\
1000                                                                                              & 1775.0650                  & 1747.6440                  \\
2000                                                                                              & 2872.2980                  & 2834.4890                  \\ \hline
\end{tabular}
\caption{Results for the agent count test. All the results are averages of 5 replications.}
\label{tab:res-acount}
\end{table}

% In the case of the number of agents test (Fig.\ref{fig:res-acount}) the behaviour of the graph fits the expectations.
In the case of the test of number of agents (Table~\ref{tab:res-acount}), the behaviour of the chart fits the expectations.
The chart increases linearly w.r.t the increase of the total number of agents (and, correspondingly, the number of PyCOMPSs tasks).
%We may also notice small convex sections that were also hypothesised although at larger scale.
Also, even if it is not clear enough from the plot, the controller time noticeably increases due to the fact that the controller has to process additional messages and handle these new agents.

% But the most interesting results come from the experiments on the number of processes (Fig.\ref{fig:res_proc}).
But the most interesting results come from the experiments on the number of processes (Table~\ref{tab:res_proc}).
For them we have two sets of columns: the left ones show the results for the default tests with a specific agent behaviour, the right ones are for the slightly modified version of it, where on each step we have added a one-second delay.
% Both graphs start at number of nodes equal to 2, and we can see that in the first case the sequential execution outperforms the distributed one.
% The reason is that the distribution of tasks and infrastructure interactions take some time.
%As we have explained in \S\,\ref{section:model-compss} due to that COMPSs is not suitable for fin-grained tasks.
We can see that, in the case without delays, there is only a slight dependency between the number of processes and the execution time.
But as soon as we imitate the harder tasks that take more time to execute, even just adding one second, the difference becomes massive as we get a clear exponential dependency.

%\begin{figure}[h]
%    \centering
%    \subfigure{\includegraphics[width=0.35\textwidth]{images/res-proc0.png}}
%    \qquad
%    \subfigure{\includegraphics[width=0.55\textwidth]{images/res-proc1.png}}
%    \caption{Results for number of processes tests. The left graph covers the case of the original agent behaviours, the right one is ones with introduced 1 second delays.}
%    \label{fig:res_proc}
%\end{figure}

% \begin{figure}[!t]
%     \begin{subfigure}[b]{0.400\textwidth}
%     \includegraphics[width=0.95\textwidth]{images/res-proc0.png}
%   %  \caption{Average wastewater volumes}
%     \label{fig:sim-water}
%     \end{subfigure}\hfill%
%     \begin{subfigure}[b]{0.600\textwidth}
%     \includegraphics[width=1\textwidth]{images/res-proc1.png}
%   % \caption{Average industry profit}
%     \label{fig:sim-budget}
%     \end{subfigure}
%     \caption{Results for number of processes tests. The left graph covers the case of the original agent behaviours, the right one is ones with introduced 1 second delays.}
%   \label{fig:res_proc}
% \end{figure}

\begin{table}[!t]
\centering
\centering
\begin{tabular}{|r|r|r|r|r|r|}
\hline
\multicolumn{1}{|c|}{\multirow{2}{*}{\begin{tabular}[c]{@{}c@{}}Number\\   of processes\end{tabular}}} & \multicolumn{1}{c|}{\multirow{2}{*}{\begin{tabular}[c]{@{}c@{}}Number\\ of nodes\end{tabular}}} & \multicolumn{2}{c|}{\begin{tabular}[c]{@{}c@{}}Time, s\\ (w/o delay)\end{tabular}} & \multicolumn{2}{c|}{\begin{tabular}[c]{@{}c@{}}Time, s\\ (with delay)\end{tabular}} \\ \cline{3-6} 
\multicolumn{1}{|c|}{}                                                                                 & \multicolumn{1}{c|}{}                                                                           & \multicolumn{1}{c|}{Tasks}               & \multicolumn{1}{c|}{Total}              & \multicolumn{1}{c|}{Tasks}               & \multicolumn{1}{c|}{Total}               \\ \hline
2                                                                                                      & 32                                                                                              & 223.4435                                 & 220.9405                                & 829.7442                                 & 826.4027                                 \\
4                                                                                                      & 64                                                                                              & 159.7769                                 & 157.7344                                & 373.3008                                 & 370.3567                                 \\
6                                                                                                      & 96                                                                                              & 161.5091                                 & 159.1821                                & 252.5065                                 & 250.0795                                 \\
8                                                                                                      & 128                                                                                             & 179.6115                                 & 176.8729                                & 194.2150                                 & 192.0736                                 \\
10                                                                                                     & 160                                                                                             & 164.2693                                 & 159.5374                                & 195.0013                                 & 191.9677                                 \\
12                                                                                                     & 192                                                                                             & 125.6789                                 & 122.7972                                & 195.0007                                 & 192.0937                                 \\
14                                                                                                     & 224                                                                                             & 128.8578                                 & 125.1992                                & 196.3293                                 & 193.1141                                 \\ \hline
\end{tabular}
\smallskip
\caption{Results for the test for the number of processes. 1 second delays were introduced to each task in the second case.}
\label{tab:res_proc}
\end{table}

Also please notice that after the number of processes exceeds the number of tasks, the changes in the computation time are almost non-existent, as COMPSs has enough resources to get all the tasks distributed with a minimal delay.

%The time results are presented as triples of total time the test took, part of that time that was spent on task computations, and part of that time that was spent by the controller (for message forwarding and action execution).

Besides the results on the tested metrics, we have obtained insights about both possible weaknesses of the system and directions for future research.
The main issue we faced was disk space usage.
As COMPSs transforms custom objects into files for transfer, this puts a strain on the data transfer infrastructure. This may result in exceeding cluster disk space quotas.
This limits the scale of the results we are able to achieve (although they still exceed the standard capabilities of the non-HPC BDI platforms).
%Data persistency via dataClay is the solution for the issue and we discuss it as future research.

%%% Local Variables:
%%% mode: latex
%%% TeX-master: "../CCIA"
%%% End:

\section{Conclusions}
This paper has presented a model for agent-based simulation in HPC and its implementation in PyCOMPSs. Our simulation model is driven by the controller, which can be effectively considered as the synchronisation point for the simulation, following the Bulk Synchronous Parallel model\cite{Valiant:1990:BMP:79173.79181}. The main advantage of our model when compared to the closest works in literature\cite{Coakley2012, Cordasco2012, Collier2012, Rubio-Campillo2014} is that our model would benefit from simulation domains that require goal-driven agents being able to perform more complex reasoning or planning. Another limitation of these frameworks (except for \cite{Coakley2012}) is that the agent communication is very limited, based on direct method calls. This kind of communication only works properly within agents executed on the same processor or in the overlapping zones, but not with agents in other processors. We are starting to test our model with a real scenario base on wastewater management of a full river basin with hundreds of pollutant producers and dozens of pollutant processors coordinating their effort to ensure pollution levels are law-compliant. 
Future research lines include other simulation problems and implementing normative reasoning by incorporating a norm monitor \cite{alvarez2016bringing} or deontic sensors \cite{Padget:2018:DS:3304415.3304483}, which would also facilitate the introduction of normative contexts.

% \jnote{but there is no discussion of or evidence for working at scale, nor does the discussion make clear the parallel execution structure with the barrier synchronisation at the controller} 

% \section{Experiments}
% \input{sections/experiments.tex}

% \begin{thebibliography}{99}
\section*{Acknowledgements}
\small
This work is partially supported by the BSC-IBM Deep Learning Center agreement, the Spanish Government through Programa Severo Ochoa (SEV-2015-0493), the 11th call on the Severo Ochoa Mobility Program in BSC, the Spanish Ministry of Science and Technology through TIN2015-65316-P project and the Generalitat de Catalunya (contract 2017-SGR-1414).
\bibliographystyle{plain}
\bibliography{references}

% \end{thebibliography}
\end{document}